\documentclass[11pt,a4paper]{article}
\usepackage{jheppub}

\usepackage{slashed}
\usepackage{youngtab}

\newcommand{\be}{\begin{equation}}
\newcommand{\ee}{\end{equation}}
\newcommand{\ba}{\begin{eqnarray}}
\newcommand{\ea}{\end{eqnarray}}

\newcommand\fverb{\setbox\fverbbox=\hbox\bgroup\verb}
\newcommand\fverbdo{\egroup\medskip\noindent%
            \fbox{\unhbox\fverbbox}\ }
\newcommand\fverbit{\egroup\item[\fbox{\unhbox\fverbbox}]}
\newbox\fverbbox

\newcommand{\nablaslash}{\not{\hbox{\kern-3pt $\nabla$}}}

\title{Cosmological singleton gravity theory and dS/LCFT correspondence}

\author{Yun~Soo~Myung}
\author{and~Taeyoon~Moon}
\affiliation{Institute of Basic Science and Department of Computer
Simulation, Inje University,\\
Gimhae 621-749, Korea}

\emailAdd{ysmyung@inje.ac.kr} \emailAdd{tymoon@inje.ac.kr}

\abstract{We study the evolution of cosmological perturbations
generated during de Sitter inflation in the singleton gravity
theory. This theory is composed of a dipole  pair in addition to
tensor. We obtain the singleton power spectra which show that the de
Sitter/logarithmic conformal field theory (dS/LCFT) correspondence
works for  computing the power spectra in the superhorizon limit.
Also we compute the  spectral indices for light singleton which
contains a logarithmic correction.}

\begin{document}
\maketitle \flushbottom



\section{Introduction}
The singleton theory is quite interesting because it provides two
coupled scalar equations which are combined to yield the degenerate
fourth-order equation which is the same equation for  the degenerate
Pais-Uhlenbeck oscillator~\cite{Pais:1950za}. The Dirac quantization
of the Pais-Uhlenbeck oscillator was carried out
in~\cite{Mannheim:2004qz,Smilga:2008pr}.  In the anti-de Sitter
(AdS) literature, this describes a dipole pair field (singleton) of
the AdS group~\cite{Flato:1986uh}. Later on, this theory was used
widely to derive the AdS/logarithmic conformal field theory (LCFT)
correspondence~\cite{Ghezelbash:1998rj,Kogan:1999bn,Myung:1999nd,Grumiller:2013at}
and the de Sitter (dS)/LCFT correspondence~\cite{Kehagias:2012pd}.
In other words, the singleton action on the AdS/dS background  is a
bulk action to derive the LCFT~\cite{Gurarie:1993xq,Flohr:2001zs} on
its boundary. Explicitly,  a dipole pair ($\varphi_1,\varphi_2$) on
AdS/dS space are
 dual to the rank-2 LCFT with two operators
($\sigma_1,\sigma_2$).

On the other hand, the  detection of  primordial gravitational waves
  by BICEP2~\cite{Ade:2014xna} has indicated that   the cosmic
inflation occurred at a high scale of $10^{16}$ GeV. A single scalar
field (inflaton)  is still known to be a  promising model for
describing the slow-roll (dS-like)
inflation~\cite{Baumann:2008aq,Baumann:2009ds}. An important issue
to be resolved indicates that  the tensor-to-scalar ratio is given
by $r=0.2^{+0.07}_{-0.05}$ (considering the dust reduction, it
reduces to $r=0.16^{+0.06}_{-0.05}$) which is outside of the $95\%$
confidence level of the Planck measurement~\cite{Ade:2013uln}.
Accordingly, many literature have provided plausible ways to reduce
the tension between BICEP2 and Planck
measurement~\cite{Hertzberg:2014aha,Choudhury:2014kma,Gong:2014cqa,Kim:2014dba,
Anchordoqui:2014bga,Bhattacharya:2014gva,Li:2014kla}. Also, it is
meaningful to mention  recent claims that the entire signal may be
due to polarized dust
emission~\cite{Mortonson:2014bja,Flauger:2014qra,Adam:2014oea}.

The dS/CFT correspondence has predicted the form of the  three-point
correlator of the operator which is dual to the inflaton
perturbation generated during slow-roll
inflation~\cite{Maldacena:2002vr}. This dual correlator was related
closely  to the three-point correlator of the curvature perturbation
generated during slow-roll inflation. Importantly, this
correspondence has provided  the first derivation of the
non-Gaussianity from the single field inflation.

  Hence, it is quite interesting to
compute the power spectrum of singleton (other than inflaton)
generated during dS inflation because its equation is a degenerate
fourth-order equation. In order to compute the power spectrum, one
needs to choose the Bunch-Davies vacuum in the subhorizon limit of
$z\to \infty$. Therefore, one has to quantize the singleton
canonically as we do for the inflaton. Also,  it is important to see
whether the dS/LCFT correspondence plays a crucial role in computing
the power spectrum in the superhorizon limit of $z\to
0$~\cite{Kehagias:2012pd}. As far as we know, there is no direct
evidence for the dS/LCFT correspondence.   We will show that the
momentum LCFT-correlators  $\langle \sigma_a(k)\sigma_b(-k)\rangle$
obtained from the extrapolation approach take the same form as the
power spectra [${\cal P}_{{ab},0}(k,-1)]\times k^{-3}$. This shows
that the dS/LCFT correspondence works well for obtaining the power
spectra in the superhorizon limit.

\section{Singleton gravity theory }

Let us first consider the singleton gravity theory where a dipole
 pair $\phi_1$ and $\phi_2$ are coupled minimally to Einstein
gravity. The  action is given by
\begin{equation} \label{SGA}
S_{\rm SG}=S_{\rm E}+S_{\rm S}=\int d^4x
\sqrt{-g}\Big[\Big(\frac{R}{2\kappa}-\Lambda\Big)-\Big(\partial_\mu\phi_1\partial^\mu\phi_2+m^2\phi_1\phi_2+\frac{\mu^2}{2}\phi_1^2
\Big)\Big],
\end{equation}
where the first two terms are introduced to provide de Sitter
background with $\Lambda>0$ and the last three terms ($S_{\rm S}$)
represent the singleton theory composed of two scalars $\phi_1$ and
$\phi_2$~\cite{Ghezelbash:1998rj,Kogan:1999bn,Myung:1999nd}. Here we
have $\kappa=8\pi G=1/M^2_{\rm P}$, $M_{\rm P}$ being the reduced
Planck mass and $m^2$ is the degenerate mass-squared for the
singleton. We stress that $S_{\rm SG}$ denotes the action for the
singleton gravity theory, whereas  $S_{\rm S}$ is the action for the
singleton theory itself.

The Einstein equation takes the form
\begin{equation} \label{ein-eq}
G_{\mu\nu} +\kappa \Lambda g_{\mu\nu}=\kappa T_{\mu\nu}
\end{equation}
with the energy-momentum tensor
\begin{equation}
T_{\mu\nu}=2\partial_{\mu}\phi_1\partial_\nu
\phi_2-g_{\mu\nu}\Big(\partial_\mu\phi_1\partial^\mu\phi_2+m^2\phi_1\phi_2+\frac{\mu^2}{2}\phi_1^2\Big).\end{equation}
On the other hand, two scalar field equations are coupled to be
\begin{equation} \label{b-eq1}
(\nabla^2-m^2)\phi_1=0,~~(\nabla^2-m^2)\phi_2=\mu^2\phi_1
\end{equation}
which are combined to give  a degenerate fourth-order   equation
\begin{equation} \label{b-eq2}
(\nabla^2-m^2)^2\phi_2=0.
\end{equation}
This reveals  the  nature of the singleton theory as $S_{\rm S}$
takes the following form upon  using  (\ref{b-eq1}) to eliminate the
auxiliary field $\phi_1$~\cite{Rivelles:2003jd,Kim:2013waf}:
\begin{equation}
S_{\rm S}=\frac{1}{2\mu^2} \int
d^4x\sqrt{-g}(\nabla^2-m^2)\phi_2(\nabla^2-m^2)\phi_2.
\end{equation}
 The solution of dS spacetime comes out when one chooses the
 vanishing scalars
 \begin{equation}
 \bar{R}=4\kappa \Lambda,~~\bar{\phi}_1=\bar{\phi}_2=0.
 \end{equation}
Explicitly,  curvature quantities are given by
\begin{equation}
\bar{R}_{\mu\nu\rho\sigma}=H^2(\bar{g}_{\mu\rho}\bar{g}_{\nu\sigma}-\bar{g}_{\mu\sigma}\bar{g}_{\nu\rho}),~~\bar{R}_{\mu\nu}=3H^2\bar{g}_{\mu\nu}
\end{equation}
with a constant Hubble parameter $H^2=\kappa \Lambda/3$. We choose
the dS background explicitly by choosing a conformal time $\eta$
\begin{eqnarray} \label{frw}
ds^2_{\rm dS}=\bar{g}_{\mu\nu}dx^\mu
dx^\nu=a(\eta)^2\Big[-d\eta^2+\delta_{ij}dx^idx^j\Big],
\end{eqnarray}
where the conformal  scale factor is
\begin{eqnarray}
a(\eta)=-\frac{1}{H\eta}\to a(t)=e^{Ht}.
\end{eqnarray}
Here the latter denotes  the scale factor with respect to cosmic
time $t$. During the dS stage, $a$ goes from small to a very large
value like $a_f/a_i\simeq 10^{30}$ which implies that the conformal
time $\eta=-1/aH(z=-k\eta)$ runs from $-\infty(\infty)$[the infinite
past] to $0^-(0)$ [the infinite future]. The two boundaries (${\rm
\partial dS}_{\infty/0}$) of dS space are located   at
$\eta=-\infty$ together with a point $\eta=0^-$ which make the
boundary compact~\cite{Maldacena:2002vr}. It is worth noting   that
the Bunch-Davies vacuum will be chosen  at $\eta=-\infty$, while the
dual (L)CFT can be thought of as living on a spatial slice at
$\eta=0^-$.

We   choose  the Newtonian gauge of $B=E=0 $ and $\bar{E}_i=0$ for
cosmological perturbation around the dS background (\ref{frw}). In
this case, the cosmologically perturbed metric can be simplified to
be
\begin{eqnarray} \label{so3-met}
ds^2=a(\eta)^2\Big[-(1+2\Psi)d\eta^2+2\Psi_i d\eta
dx^{i}+\Big\{(1+2\Phi)\delta_{ij}+h_{ij}\Big\}dx^idx^j\Big]
\end{eqnarray}
with transverse-traceless tensor $\partial_ih^{ij}=h=0$. Also, one
has the scalar perturbations \begin{equation} \phi_1=
\bar{\phi}_1+\varphi_1,~~\phi_2= \bar{\phi}_2+\varphi_2.
\end{equation}
In order to get the cosmological perturbed equations, one linearize
the Einstein equation (\ref{ein-eq})  directly around the dS
\begin{eqnarray}
\delta R_{\mu\nu}(h)-3H^2h_{\mu\nu}=0 \to
\bar{\nabla}^2h_{ij}=0.\label{heq}
\end{eqnarray}
We would like to mention briefly two metric scalars $\Psi$ and
$\Phi$, and a vector $\Psi_i$. The linearized Einstein  equation
requires $\Psi=-\Phi$ which was  used to define the comoving
curvature perturbation in the slow-roll inflation and thus, they are
not physically propagating modes. In the dS inflation, there is no
coupling between $\{\Psi,\Phi\}$ and $\{\varphi_1,\varphi_2\}$
because of $\bar{\phi}_1=\bar{\phi}_2=0$. The vector is also a
non-propagating mode in the singleton gravity theory because it has
no its kinetic term.  The linearized scalar equations are given by
\begin{eqnarray}
&&(\bar{\nabla}^2-m^2)\varphi_1=0,\nonumber\\
&&\label{sing-eq1}(\bar{\nabla}^2-m^2)\varphi_2=\mu^2\varphi_1.
\end{eqnarray}
These are combined to provide a degenerate fourth-order scalar
equation
\begin{equation} \label{sing-eq2}
(\bar{\nabla}^2-m^2)^2\varphi_2=0,
\end{equation}
which is our main equation to be solved for cosmological purpose.

\section{dS/LCFT correspondence in the superhorizon}
First of all, we briefly review what are similarities and
differences between AdS/CFT and dS/CFT dictionaries. The first
version of the AdS/CFT dictionary was stated in terms of an
equivalence between bulk and boundary partition functions in the
presence of deformations:
\begin{equation}
Z_{\rm bulk}[\phi_0,{\cal M}]=Z_{\rm CFT}[\phi_0,{\cal O},\partial
{\cal M}],
\end{equation} where on the bulk side $\phi_0$  specifies the boundary conditions of bulk field $\phi$ propagating on
${\cal M}$,  whereas  on the boundary CFT $\phi_0$  denotes the
sources of operators ${\cal O}$ on the boundary $\partial {\cal M}$.
Correlator of dual CFT can be computed by differentiating the
partition function with respect to the sources and then, setting
them to zero as
\begin{equation} \label{diff-c}
\langle {\cal O}({\bf x}){\cal O}({\bf y})\rangle_{\rm
d}=\frac{\delta^2 Z_{\rm CFT}}{\delta\phi_0({\bf
x})\delta\phi_0({\bf y})}\Big|_{\phi_0=0}.
\end{equation}
This is called ``differentiate" (GKPW)
dictionary~\cite{Banks:1998dd}. The second version consists of
computing bulk-to-boundary propagators first and pulling  CFT
correlators to the boundary  as
\begin{equation}\label{extra-c}
\langle {\cal O}({\bf x}){\cal O}({\bf y})\rangle_{\rm
e}=\lim_{z\to
0}z^{-2\Delta}\langle \phi({\bf x},z)\phi({\bf y},z)\rangle.
\end{equation} 
This version was used
in~\cite{Susskind:1998dq} and was referred to ``extrapolate" (BDHM)
dictionary~\cite{Polchinski:1999ry}.

Concerning correlation functions of a free massive scalar in AdS and
dS, the following three statements appear importantly~\cite{Harlow:2011ke}:\\
 (a) In Euclidean  AdS$_{d+1}$ with $\ell^2_{\rm AdS}=1$, either the differentiation of
 the partition function with respect to sources or extrapolation of
 the bulk operators to the boundary produce CFT correlators of an operator with dimension
 $\Delta=\frac{d}{2}+\frac{\sqrt{d^2+4m^2}}{2}$. \\
 (b) In Lorentzian dS$_{d+1}$ with $\ell^2_{\rm dS}=1$, the extrapolated bulk
 correlators  are  a sum of two contributions. One is the leading
 behavior of a CFT correlator of an operator with dimension
 $d-\delta=\frac{d}{2}-\frac{\sqrt{d^2-4m^2}}{2}$, whereas the other
comes from the leading
 behavior of a CFT correlator of an operator with dimension
 $\delta=\frac{d}{2}+\frac{\sqrt{d^2-4m^2}}{2}$. \\
(c) In Lorentzian dS$_{d+1}$ with $\ell^2_{\rm dS}=1$, functional
derivatives of late-time Schr\"{o}dinger wave-function produce CFT
correlators with dimension
$\delta$ only. \\
The dominant term in (b) was computed by Witten for a particular
scalar~\cite{Witten:2001kn},  while a massless version of statement
(c) was firstly made by Maldacena~\cite{Maldacena:2002vr}. This
implies that the dS/CFT ``extrapolate" and ``differentiate"
dictionaries are inequivalent to each other. Particularly, the
dimension of  CFT operators associated to a massive scalar is
different:
$\triangle_+(=\delta)=\frac{3}{2}+\sqrt{\frac{9}{4}-\frac{m^2}{H^2}}$
for ``differentiate" dictionary and both
$\triangle_{\pm}=\frac{3}{2}\pm
\sqrt{\frac{9}{4}-\frac{m^2}{H^2}}(\triangle_-=w)$ for
``extrapolate" dictionary in four dimensional dS space. Accordingly,
following  (c) to compute cosmological correlator of a massive
scalar, it in momentum space is inversely proportional to  CFT
correlator with dimension $\Delta_+$ as
\begin{equation}
 \langle\phi(k)\phi(-k)\rangle \propto \frac{1}{2{\rm Re}\langle {\cal O}(k){\cal
 O}(-k)\rangle}_{\rm d}\propto \frac{1}{k^{-3+2\triangle_+}}=k^{2w-3},\end{equation}
which leads to the power spectrum for a massive scalar in the
superhorizon limit.  If one employs (c) to derive the dS/LCFT
correspondence, the approach (c) may break down for deriving LCFT
correlators because all LCFT correlators in AdS$_{d+1}$ were derived
based on the extrapolation approach
(b)~\cite{Ghezelbash:1998rj,Kogan:1999bn,Myung:1999nd,Grumiller:2013at}.
Hence, we wish to use the extrapolation approach (b) to derive the
LCFT correlators from the bulk correlators. In this case, the
cosmological correlator is directly proportional to the CFT
correlator with different dimension $\triangle_-$
\begin{equation} \label{dir-rel}
\langle\phi(k)\phi(-k)\rangle \propto
\langle\sigma(k)\sigma(-k)\rangle_{\rm e} \propto k^{2w-3}
\end{equation}
 as was shown in
(\ref{extra-c}).

 To develop the dS/LCFT
correspondence~\cite{Kehagias:2012pd}, we first solve
Eqs.(\ref{sing-eq1}) and (\ref{sing-eq2}) for  the singleton gravity
theory in the superhorizon limit of $\eta\to 0^-$. Their solutions
are given by
\begin{equation}
\varphi_{1,0} \sim \eta^w,~~\varphi_{2,0} \sim \eta^w\ln[-\eta]
\end{equation}
with
\begin{equation}
w=\frac{3}{2}\Bigg(1-\sqrt{1-\frac{4m^2}{9H^2}}\Bigg).
\end{equation}
The scaling of $\varphi_{a,0}$ with $a=1,2$ is not conventional as
they transform under
\begin{equation} \label{trans}
\varphi_{1,0} \to \lambda^w\varphi_{1,0},~~\varphi_{2,0} \to
\lambda^w\Big[\varphi_{2,0}+\ln (\lambda)\varphi_{1,0}\Big].
\end{equation}
 A pair of dipole  fields $(\varphi_1,\varphi_2)$ is  coupled
to
 $(\sigma_1,\sigma_2)$-operators on the boundary (${\rm \partial dS}$) of $\eta\to 0^-$.
 The explicit connection between $\varphi_{a,0}$ and
$\sigma_a$ is encoded by~\cite{Seery:2006tq}
\begin{eqnarray} \label{ds-lcft}
&&Z_{S}[{\varphi_{a,0}}]=Z_{\rm LCFT}[{\varphi_{a,0}}],\\
\label{ds-lcft1}&&Z_{S}[{\varphi_{a,0}}]=e^{-\delta S_{\rm
S}[\{\varphi_{a,0}\}]},\\
\label{ds-lcft2}&&Z_{\rm LCFT}[\varphi_{a,0}]= \langle
e^{-\int_{{\rm
\partial dS}_0} d^3x\varphi_{a,0}({\bf x})\sigma_a({\bf
x})}\rangle,
\end{eqnarray}
where the expectation value $\langle \cdots \rangle$  is taken in
the LCFT with the boundary fields $\varphi_{a,0}$  as  sources.
Eq.(\ref{ds-lcft}) is  a statement of the dS/LCFT correspondence.
Here the bulk action is given by
\begin{equation}
\delta S_{\rm S}[\{\varphi_a\}]=-\int_{\rm dS}
d^4x\sqrt{-\bar{g}}\Big[\partial_\mu\varphi_1\partial^\mu\varphi_2+m^2\varphi_1\varphi_2+\frac{\mu^2}{2}\varphi_1^2\Big].
\end{equation}
 The bulk transformation (\ref{trans}) indicates that two operator
 $\sigma_a$ of conformal dimension $w$ transform under dilations as
\begin{equation}
i[D,\sigma_a]=\Big(x^i\partial_i \delta^b_a+\Delta^b_a\Big)\sigma_b,
\end{equation}
where a dimension matrix $\Delta^b_a$ is brought to the Jordan cell
form as
\begin{equation} \label{b-jcell}
\Delta^b_a=
                      \left(
                        \begin{array}{cc}
                          w & 0 \\
                          1 & w \\
                        \end{array}
                      \right).
\end{equation}
This implies that $\sigma_a$ transform under dilations of ${\bf x}
\to \lambda {\bf x}$ as
\begin{equation}
\sigma_a({\bf x}) \to \sigma'_a(\lambda {\bf x})=\Big(e^{\Delta \ln
\lambda}\Big)^b_a\sigma_b(\lambda{\bf x}).
\end{equation}

In order to find the LCFT correlators $\langle \sigma_a({\bf
x})\sigma_b({\bf y})\rangle$, one might use the Ward identities for
scale and special conformal transformations~\cite{Kehagias:2012pd}.
In this work, we wish to rederive them by using the extrapolation
approach (b) (see Appendix for detail computations). The two-point
functions of $\sigma_1$ and $\sigma_2$ are determined by
\begin{eqnarray}
&& \label{cc0} _{\rm C}\langle \sigma_1({\bf x})\sigma_1({\bf
y})\rangle_{\rm C}=0,\\
\label{cc1} && _{\rm C}\langle \sigma_1({\bf x})\sigma_2({\bf
y})\rangle_{\rm C}=~_{\rm C}\langle \sigma_2({\bf x})\sigma_1({\bf
y})\rangle_{\rm C}=\frac{A}{|{\bf x}-{\bf
y}|^{2w}}, \\
&&  \label{cc2}_{\rm C}\langle \sigma_2({\bf x})\sigma_2({\bf
y})\rangle_{\rm C}=\frac{A}{|{\bf x}-{\bf y}|^{2w}}\Big(-2\ln|{\bf
x}-{\bf y}|+D\Big).
\end{eqnarray}
Here $w$ is a degenerate dimension of $\sigma_1$ and $\sigma_2$. The
coefficient $A=w(2w-3)$ is determined by the normalization of
$\sigma_1$ and $\sigma_2$. However, $D$ is arbitrary.  The CFT
vacuum $|0\rangle_{\rm C}$ is defined by three Virasoro operators
$L_n|0\rangle_{\rm C}=0$ for $n=0,\pm1$. The highest-weight state
$|\sigma_a\rangle_{\rm C}=\sigma_a(0)|0\rangle_{\rm C}$ for two
primary fields $\sigma_a$ of conformal weight $h=w/2$ is defined by
\begin{equation} \label{cft-jcell}
L_0|\sigma_1\rangle_{\rm C}=h|\sigma_1\rangle_{\rm C},~~
L_0|\sigma_2\rangle_{\rm C}=|\sigma_1\rangle_{\rm C}+
h|\sigma_2\rangle_{\rm C},~~L_n|\sigma_a\rangle_{\rm C}=0 ~{\rm
for}~n>0.
\end{equation}
This implies that for any pair of degenerate operators $\sigma_1$
and $\sigma_2$ (logarithmic pair),  the Hamiltonian ($L_0$) becomes
non-diagonalizable which shows us  a crucial difference from an
ordinary  CFT. Actually, Eq.(\ref{cft-jcell}) represents the CFT
version of the bulk transformation (\ref{trans}).
Eqs.(\ref{cc0})-(\ref{cc2}) are summarized to be
\begin{equation} \label{lcft-mat}
_{\rm C}\langle \sigma_a({\bf x})\sigma_b({\bf y})\rangle_{\rm C}
=\left(
   \begin{array}{cc}
     0 & {\rm CFT} \\
     {\rm CFT} & {\rm LCFT} \\
   \end{array}
 \right),
 \end{equation}
where CFT and LCFT represent their correlators in (\ref{cc1}) and
(\ref{cc2}), respectively.

 In order to
derive  the relevant correlators in momentum space, one has to use
the relation
\begin{equation}
\frac{1}{|{\bf x}-{\bf y}|^{2w}}=\frac{\Gamma(\frac{3}{2}-w)}{4^w
\pi^{3/2}\Gamma(w)}\int d^3{\bf k}|{\bf k}|^{2w-3}e^{i{\bf k}\cdot
({\bf x}-{\bf y})},\end{equation} where we observe an
inverse-relation of exponent $2w$ between $|{\bf x}|$-space and
$k=|{\bf k}|$-space. Finally, the correlators in momentum space are
easily evaluated as~\cite{Kehagias:2012pd}
\begin{eqnarray}
 \label{m0} \langle \sigma_1({\bf k}_1)\sigma_1({\bf
k}_2)\rangle'&=&0,\\
\label{m1} \langle \sigma_1({\bf k}_1)\sigma_2({\bf
k}_2)\rangle'&=&\frac{A_0(w)}{k_1^{3-2w}}, \\
 \langle \sigma_2({\bf k}_1)\sigma_2({\bf k}_2)\rangle'&=&D\langle
\sigma_1({\bf k}_1)\sigma_2({\bf
k}_2)\rangle'+\frac{\partial}{\partial w}\langle \sigma_1({\bf
k}_1)\sigma_2({\bf k}_2)\rangle' \nonumber \\
 \label{m2}&=&\frac{A_0(w)}{k_1^{3-2w}}\Bigg(2\ln[k_1]+D+\frac{A_{0,w}}{A_0(w)}\Bigg),
\end{eqnarray}
where the prime ($'$) represents correlators without the
$(2\pi)^3\delta^3(\Sigma_i{\bf k}_i)$ and $A_{0,w}=4w-3$ denotes
derivatives of $A_0(w)=w(2w-3)$ with respect to $w$. These
correlators will be compared to the power spectra in the
superhorizon limit of $z\to 0$.

\section{Singleton propagation in dS spacetime}

In order to compute the power spectrum, we have to know the solution
to singleton equations Eqs.(\ref{sing-eq1}) and (\ref{sing-eq2}) in
the  whole range of $\eta(z)$. For this purpose, the scalars
$\varphi_{i}$ can be expanded in Fourier modes $\phi^{i}_{\bf
k}(\eta)$
\begin{eqnarray}\label{scafou}
\varphi_{i}(\eta,{\bf x})=\frac{1}{(2\pi)^{\frac{3}{2}}}\int
d^3k~\phi^{i}_{\bf k}(\eta)e^{i{\bf k}\cdot{\bf x}}.
\end{eqnarray}
The first equation of (\ref{sing-eq1}) leads to
\begin{eqnarray}\label{scalar-eq2}
\Bigg[\frac{d^2}{d \eta^2}-\frac{2}{\eta}\frac{d}{d
\eta}+k^2+\frac{m^2}{H^2}\frac{1}{\eta^2}\Bigg]\phi^1_{\bf
k}(\eta)=0,
\end{eqnarray}
which can be further transformed into
\begin{eqnarray}\label{scalar-eq3}
\Bigg[\frac{d^2}{d\eta^2}+k^2-\frac{2}{\eta^2}+\frac{m^2}{H^2}\frac{1}{\eta^2}\Bigg]\tilde{\phi}^1_{\bf
k}(\eta)=0
\end{eqnarray}
for $\tilde{\phi}^1_{\bf k}=a\phi^1_{\bf k}=-\phi^1_{\bf
k}/(H\eta)=\frac{k}{Hz}\phi^1_{\bf k}$. Expressing
(\ref{scalar-eq3}) in terms of $z=-k\eta$ leads  to
\begin{eqnarray}\label{scalars-eq4}
\Bigg[\frac{d^2}{dz^2}+1-\Big(2-\frac{m^2}{H^2}\Big)\frac{1}{z^2}\Bigg]\tilde{\phi}^1_{\bf
k}(z)=0.
\end{eqnarray}
Introducing $\tilde{\phi}^1_{\bf
k}=\sqrt{z}\tilde{\tilde{\phi}}^1_{\bf k}$ further,  it  leads to
the Bessel's equation as
\begin{eqnarray}\label{scalar-eq4}
\Bigg[\frac{d^2}{dz^2}+\frac{1}{z}\frac{d}{dz}+1-\frac{\nu^2}{z^2}\Bigg]\tilde{\tilde{\phi}}^1_{\bf
k}(z)=0
\end{eqnarray}
with the index
\begin{equation}
\nu=\sqrt{\frac{9}{4}-\frac{m^2}{H^2}}.
\end{equation}
The solution to (\ref{scalar-eq4}) is given by the Hankel function
$H^{(1)}_\nu$. Accordingly, one has the solution to
(\ref{scalar-eq2})
\begin{equation} \label{scalar-eq5}
\phi^1_{\bf k}(z)={\cal
C}\frac{\sqrt{z}}{a}\tilde{\tilde{\phi}}^1_{\bf k}={\cal
C}\frac{H}{k}z^{3/2}H^{(1)}_{\nu}(z)
\end{equation}
with ${\cal C}$ undetermined constant.  In the subhorizon limit of
$z\to \infty$, Eq.(\ref{scalar-eq2}) reduces to
\begin{equation}\label{scalar-eq6}
\Big[\frac{d^2}{dz^2}-\frac{2}{z}\frac{d}{dz}+1\Big]\phi^{1}_{{\bf
k},\infty}(z)=0
\end{equation}
which leads the positive-frequency  solution with the normalization
$1/\sqrt{2k}$
\begin{equation} \label{scalar-eq7}
\phi^{1}_{{\bf k},\infty}(z)=\frac{H}{\sqrt{2k^3}}(i+z)e^{iz}.
\end{equation}
This is a typical mode solution of a massless scalar propagating on
dS spacetime.  Inspired by (\ref{scalar-eq7}) and asymptotic form of
$H^{(1)}_\nu$, $\phi^1_{\bf k}(z)$ is fixed  by
\begin{equation} \label{scalar-eq10}
\phi^1_{\bf k}(z)=\frac{H}{\sqrt{2k^3}}
\sqrt{\frac{\pi}{2}}e^{i(\frac{\pi\nu}{2}+\frac{\pi}{4})}z^{3/2}H^{(1)}_{\nu}(z).
\end{equation}

In the superhorizon limit of $z\to0$, Eq.(\ref{scalar-eq2}) takes
the form
\begin{equation}\label{scalar-eq11}
\Bigg[\frac{d^2}{dz^2}-\frac{2}{z}\frac{d}{dz}+\frac{m^2}{H^2}\frac{1}{z^2}\Bigg]\phi^1_{{\bf
k},0}(z)=0,
\end{equation}
whose solution is
\begin{equation}
\phi^1_{{\bf k},0}(z)=\frac{H}{\sqrt{2k^3}}z^{w}
\end{equation}
with
\begin{equation}
w=\frac{3}{2}-\nu.
\end{equation}

On the other hand,  plugging (\ref{scafou}) into (\ref{sing-eq2})
leads to the degenerate fourth-order differential equation
\begin{eqnarray}
\Bigg[\eta^2\frac{d^2}{d\eta^2}-2\eta\frac{d}{d\eta}+k^2\eta^2+\frac{m^2}{H^2}\Bigg]^2\phi^2_{\bf
k}(\eta)=0\label{s2-eq2}
\end{eqnarray}
which seems difficult to be solved directly. However, we may solve
Eq.(\ref{s2-eq2}) in  the two limits of subhorizon and superhorizon.
In the subhorizon limit of $z\to \infty$, Eq.(\ref{s2-eq2}) takes
the form
\begin{equation}\label{sub-eq1}
\Bigg[\frac{d^4}{dz^4}+2\Big(1-\frac{1}{z^2}\Big)\frac{d^2}{dz^2}+\frac{4}{z^3}\frac{d}{dz}+\Big(1-\frac{2}{z^2}\Big)\Bigg]\phi^2_{{\bf
k},\infty}=0.
\end{equation}
whose direct solution is given by
\begin{eqnarray} \label{sub-sol}
\phi^{2,d}_{{\bf
k},\infty}=\Big[\tilde{c}_2(i+z)+\tilde{c}_1\Big(2i+(z-i)e^{-2iz}{\rm
Ei}(2iz)\Big)\Big]e^{iz}
\end{eqnarray}
with two coefficients $\tilde{c}_1$ and $\tilde{c}_2$. The c.c. of
$\phi^{2,d}_{{\bf k},\infty}$ is a solution to (\ref{sub-eq1}) too.
Here ${\rm Ei}(2iz)$ is the exponential integral function defined
by~\cite{AS}
\begin{eqnarray}
{\rm Ei}(2iz)={\rm Ci}(2z)+i{\rm Si}(2z)+i\frac{\pi}{2},
\end{eqnarray}
where the cosine-integral and sine-integral functions are given by
\begin{eqnarray}
{\rm Ci}(2z)=\int^{2z}_{0}\frac{{\rm cos} t}{t}dt,~~{\rm
Si}(2z)=\int^{2z}_{0}\frac{{\rm sin} t}{t}dt.
\end{eqnarray}
We note that ${\rm Ei}(2iz)$ satisfies the fourth-order equation
\begin{eqnarray} (z-i)z^3\frac{d^4{\rm Ei}}{dz^4}&-&4iz^4 \frac{d^3{\rm Ei}}{dz^3}+2z(i-z-4iz^2-2z^3)\frac{d^2{\rm Ei}}{dz^2}\nonumber
\\
&-&4(i-z-iz^2+2z^3)\frac{d{\rm Ei}}{dz}-8e^{2iz}=0.
\end{eqnarray}
However, we wish to point out that the direct solution
(\ref{sub-sol}) is not suitable for choosing the Bunch-Davies vacuum
to give quantum fluctuations. In order to find an appropriate
solution, we note that
$(\bar{\nabla}^2-m^2)\varphi_2=\mu^2\varphi_1$ in (\ref{sing-eq1})
reduces to  in the subhorizon limit
\begin{equation} \label{phi2t}
\Big[\frac{d^2}{dz^2}-\frac{2}{z}\frac{d}{dz}+1\Big]\phi^{2}_{{\bf
k},\infty}(z)=0,
\end{equation}
whose solution is \begin{equation} \label{phi2s-sol} \phi^{2}_{{\bf
k},\infty}(z)=\tilde{c}_2(i+z)e^{iz}.
\end{equation}
We note that $ \phi^{2}_{{\bf k},\infty}(z)$ is included as the
first term of (\ref{sub-sol}) [as a solution to the fourth-order
equation (\ref{sub-eq1})].

 On the other
hand, Eq.(\ref{s2-eq2}) takes the form in the superhorizon limit of
$z\to 0$ as
\begin{eqnarray}
\Bigg[z^2\frac{d^2}{dz^2}-2z\frac{d}{dz}+\frac{m^2}{H^2}\Bigg]^2\phi^2_{{\bf
k},0}(z)=0\label{super-eq2}
\end{eqnarray}
whose solution is given by
\begin{equation}\label{super-phi2}
\phi^2_{{\bf k},0}(z)\propto z^w\ln z.
\end{equation}
This also satisfies
\begin{eqnarray}
(-H^2)\Bigg[z^2\frac{d^2}{dz^2}-2z\frac{d}{dz}+\frac{m^2}{H^2}\Bigg]\phi^2_{{\bf
k},0}(z)=\mu^2\phi^1_{{\bf k},0}(z)\label{super-eq3}
\end{eqnarray}
 for $\mu^2=(3-2w)H^2$ which is the superhorizon limit of
Eq.(\ref{sing-eq1}).
  The presence of ``$\ln z$" implies that
(\ref{super-phi2}) is a solution to the fourth-order equation
(\ref{super-eq2})

Finally, the trick used in~\cite{Kogan:1999bn} implies that  one may
solve  (\ref{s2-eq2}) directly by  differentiating
$(\bar{\nabla}^2-m^2)\varphi_1=0$  with respect to $m^2$. The
explicit steps are  given by
\begin{eqnarray}
\frac{d}{dm^2}&\times&\left(-z^2H^2\frac{d^2}{dz^2}+2z
H^2\frac{d}{dz}-z^2H^2-m^2\right)\phi_{\bf k}^1(z)
=0\\
&\rightarrow&\left(-z^2H^2\frac{d^2}{dz^2}+2z
H^2\frac{d}{dz}-z^2H^2-m^2\right)\frac{d}{dm^2}\phi_{\bf
k}^1(z) =\phi_{\bf k}^1(z)\\
&\leftrightarrow&\left(-z^2H^2\frac{d^2}{dz^2}+2z
H^2\frac{d}{dz}-z^2H^2-m^2\right)\phi_{\bf k}^2(z) =\mu^2\phi_{\bf
k}^1(z)\label{phi2e1}
\end{eqnarray}
which provides   a way  to obtain $\phi_{\bf k}^2(z)$ from
$\phi_{\bf k}^1(z)$ as
\begin{equation} \phi_{\bf k}^2(z)=\mu^2\frac{d}{dm^2}\phi_{\bf
k}^1(z)\label{phi2e}.
\end{equation}
We note that  (\ref{s2-eq2}) can be obtained by  acting
$(\bar{\nabla}^2-m^2)$ on  (\ref{phi2e1}).   Explicitly,
$\frac{d}{dm^2}\phi_{\bf k}^1(z)$ is computed to be
\begin{eqnarray}
\frac{d}{dm^2}\phi_{\bf k}^1(z)&=&-\frac{1}{2\nu
H\sqrt{2k^3}}\sqrt{\frac{\pi}{2}}e^{i\left(\frac{\pi\nu}{2}+\frac{\pi}{4}\right)}z^{3/2}
\Bigg\{\pi\Big(\frac{i}{2}-\cot[\nu\pi]\Big)H_{\nu}^{(1)}
+i\csc[\nu\pi]\times\nonumber\\
&&\hspace*{10em}\Big(e^{-\nu\pi
i}\frac{\partial}{\partial\nu}J_{\nu}-\frac{\partial}{\partial\nu}J_{-\nu}
-\pi i e^{-\nu\pi i}J_{\nu}\Big)\Bigg\},\label{phi1e}
\end{eqnarray}
where
\begin{eqnarray}
\frac{\partial}{\partial\nu}J_{\nu}(z)=J_{\nu}\ln\Big[\frac{z}{2}\Big]
-\Big(\frac{z}{2}\Big)^{\nu}\sum_{k=0}^{\infty}(-1)^{k}
\frac{\psi(\nu+k+1)}{\Gamma(\nu+k+1)}\frac{(\frac{z^2}{4})^k}{k!}
\end{eqnarray}
with the digamma function $\psi(x)=\partial\ln[\Gamma(x)]/\partial
x$. Here we observe the appearance of $\ln[z]$-term. It turns out
that  $\phi_{\bf k}^2(z)$ takes the form when considering
$J_{\pm\nu} \to \Gamma(\pm\nu+1)^{-1}(z/2)^{\pm\nu}$ in the
superhorizon limit of $z\to0$ as
\begin{eqnarray}
\phi_{\bf k}^2(z)\sim z^{w}\ln[z],
\end{eqnarray}
which recovers (\ref{super-phi2}).  We mention that
$\frac{\partial}{\partial\nu}J_{-\nu}$ in (\ref{phi1e}) is dominant
because it behaves as $z^{-\nu}\ln[z]$ in the superhorizon limit of
$z\to 0$. However, we  do not recover its asymptotic form
(\ref{phi2s-sol}) in the subhorizon limit of $z\to\infty$. Hence, it
is not easy  to obtain a full solution $\phi_{\bf k}^2(z)$ to
(\ref{s2-eq2}) by the trick used in~\cite{Kogan:1999bn}.
Fortunately, its superhorizon-limit solution (\ref{super-phi2})
could be found by this trick.

\section{Power spectra}

The power spectrum is  defined by the two-point  function which
could be computed when one chooses   the Bunch-Davies (BD) vacuum
state $|0\rangle_{\rm BD}$ in the subhorizon limit (${\rm \partial
dS}_\infty$)  of $\eta\to -\infty(z\to
\infty)$~\cite{Baumann:2009ds}. The defining relation is given by
\begin{equation}
_{\rm BD}\langle0|{\cal F}(\eta,\bold{x}){\cal
F}(\eta,\bold{y})|0\rangle_{\rm BD}=\int d^3k \frac{{\cal P}_{\cal
F}}{4\pi k^3}e^{i \bold{k}\cdot (\bold{x}-\bold{y})},
\end{equation}
where ${\cal F}$ represents  singleton  and tensor and
$k=\sqrt{\bold{k}\cdot \bold{k}}$ is the comoving wave number.
Quantum fluctuations were created on all length scales with wave
number $k$. Cosmologically relevant fluctuations start their lives
inside the Hubble radius which defines the subhorizon: $k~\gg aH$.
On later, the comoving Hubble radius $1/(aH)$ shrinks during
inflation while keeping the wavenumber $k$ constant. Eventually, all
fluctuations exit the comoving Hubble radius, they reside on  the
superhorizon region of $k~\ll aH$ after horizon crossing.

In general, one may compute the power spectrum of scalar and tensor
by taking the BD vacuum.
 In the dS inflation, we choose the subhorizon limit
of  $z\to \infty$  to define the BD vacuum. This implies that in the
infinite past of $\eta\to -\infty(z\to \infty)$, all observable
modes had time-independent frequencies $\omega=k$ and the
Mukhanov-Sasaki equation reduces to ${\cal F}''_{{\bf
k},\infty}+k^2{\cal F}_{{\bf k},\infty}\approx0$ whose positive
solution is given by ${\cal F}_{{\bf
k},\infty}=e^{-ik\eta}/\sqrt{2k}=e^{iz}/\sqrt{2k}$. This defines a
preferable set of mode functions and a unique physical vacuum, the
BD vacuum $|0\rangle_{\rm BD}$.

On the other hand,  we choose the superhorizon region  of $z \ll 1$
to get a finite form of the power spectrum which stays alive after
decaying. For example, fluctuations of a massless scalar
($\bar{\nabla}^2\delta \phi=0$) and tensor
($\bar{\nabla}^2h_{ij}=0$) with different normalization originate on
subhorizon scales and they propagate for a long time on superhorizon
scales. This can be checked by computing their power spectra given
by
\begin{eqnarray} \label{powerst}
{\cal P}_{\rm \delta\phi}&=&\frac{H^2}{(2\pi)^2}[1+z^2],\\
 \label{powerst1}{\cal P}_{\rm h}&=&2\times \Big(\frac{2}{M_{\rm
P}}\Big)^2\times{\cal P}_{\rm \delta\phi}= \frac{2H^2}{\pi^2M^2_{\rm
P}}[1+z^2].
\end{eqnarray}
In the limit of $z\to 0$, they are finite as
\begin{equation}\label{fpowerst} {\cal P}_{{\rm
\delta \phi},0}=\frac{H^2}{(2\pi)^2},~~{\cal P}_{{\rm h},0}=
\frac{2H^2}{\pi^2M^2_{\rm P}}.
\end{equation}
Accordingly, it would be very interesting  to check what happens
when one computes the power spectra for the dipole  pair (singleton)
generated from during the dS inflation in the framework of the
singleton gravity theory.

 To compute the power spectrum, we have to know the commutation relations and the Wronskian conditions. The canonical
conjugate momenta are given by
\begin{equation}
\pi_1=a^2\frac{d\varphi_2}{d\eta},~~\pi_2=a^2\frac{d\varphi_1}{d\eta}.
\end{equation}
The canonical quantization is accomplished by imposing equal-time
commutation relations:
\begin{eqnarray}\label{comm}
[\hat{\varphi}_{1}(\eta,{\bf x}),\hat{\pi}_{1}(\eta,{\bf
y})]=i\delta^3({\bf x}-{\bf y}),~~[\hat{\varphi}_2(\eta,{\bf
x}),\hat{\pi}_{2}(\eta,{\bf y})]=i\delta^3({\bf x}-{\bf y}).
\end{eqnarray}
 The two operators $\hat{\varphi}_{1}$ and
$\hat{\varphi}_{2}$ are expanded in terms of Fourier modes
as~\cite{Rivelles:2003jd,Jimenez:2012ak,Kim:2013waf}
\begin{eqnarray}\label{hex1}
\hat{\varphi}_{1}(z,{\bf x})&=&\frac{1}{(2\pi)^{\frac{3}{2}}}\int
d^3kN\Bigg[\Big(i\hat{c}_1({\bf k})\phi^1_{\bf k}(z)e^{i{\bf
k}\cdot{\bf
x}}\Big)+{\rm h.c.}\Bigg], \\
\label{hex2} \hat{\varphi}_2(z,{\bf
x})&=&\frac{1}{(2\pi)^{\frac{3}{2}}}\int
d^3k\tilde{N}\Bigg[\Big(\hat{c}_2({\bf k})\phi^1_{\rm
k}(z)+\hat{c}_1({\bf k})\phi^2_{\rm k}(z)\Big)e^{i{\bf k}\cdot{\bf
x}}+{\rm h.c.}\Bigg]
\end{eqnarray}
with $N$ and $\tilde{N}$ the normalization constants.  Plugging
(\ref{hex1}) and (\ref{hex2}) into (\ref{comm}) determines the
relation of normalization constants as $N\tilde{N}=1/2k $ and
commutation relations between $\hat{c}_a({\bf k})$ and
$\hat{c}^{\dagger}_b({\bf k}')$ as
 \begin{equation} \label{scft}
 [\hat{c}_a({\bf k}), \hat{c}^{\dagger}_b({\bf k}')]= 2k
 \left(
  \begin{array}{cc}
   0 & -i  \\
    i & 1 \\
  \end{array}
 \right)\delta^3({\bf k}-{\bf k}')
 \end{equation}
 which reflects the quantization of singleton.
Here, the commutation relation of $[\hat{c}_2({\bf k}),
\hat{c}^{\dagger}_2({\bf k}')]$ is implemented  by the following
Wronskian condition with (\ref{scalar-eq7}) and
$\tilde{c}_2=-iH/(2\sqrt{2k^3})$ in (\ref{phi2s-sol}):
\begin{eqnarray}
a^2\Big(\phi^1_{{\bf k},\infty}\frac{d\phi^{2*}_{{\bf
k},\infty}}{dz}-\phi^{2*}_{{\bf k},\infty}\frac{d\phi^{1}_{{\rm
k},\infty}}{dz}+\phi^{1*}_{{\bf k},\infty}\frac{d\phi^{2}_{{\bf
k},\infty}}{dz}-\phi^{2}_{{\bf k},\infty}\frac{d\phi^{1*}_{{\rm
k},\infty}}{dz}\Big)=\frac{1}{k}.
\end{eqnarray}
It is important to note that the commutation relations (\ref{scft})
were used to
 derive the power spectra of conformal gravity~\cite{Myung:2014cra}.
On the other hand, if one uses the  solution $\phi^{1}_{{\bf
k},\infty}$ (\ref{scalar-eq7}) and $\phi^{2,d}_{{\bf k},\infty}$
(\ref{sub-sol}), the Wronskian condition  leads to
\begin{eqnarray}
&&a^2\Big(\phi^1_{{\bf k},\infty}\frac{d\phi^{2,d*}_{{\bf
k},\infty}}{dz}-\phi^{2,d*}_{{\bf k},\infty}\frac{d\phi^{1}_{{\rm
k},\infty}}{dz}+\phi^{1*}_{{\bf k},\infty}\frac{d\phi^{2,d}_{{\bf
k},\infty}}{dz}-\phi^{2,d}_{{\bf k},\infty}\frac{d\phi^{1*}_{{\rm
k},\infty}}{dz}\Big)\nonumber
\\
&&=-\sqrt{\frac{k}{2}}\frac{1}{H}\Bigg[2i(-\tilde{c}_2+\tilde{c}^*_2)+(\tilde{c}_1+\tilde{c}_1^*)\Big(\frac{1}{z^3}+\frac{3}{z}\Big)\Bigg]
\end{eqnarray}
which cannot be independent of $z$ unless
$\tilde{c}_1=\tilde{c}_1^*=0$, This explains  why the direct
solution $\phi^{2,d}_{{\bf k},\infty}$ (\ref{sub-sol}) is not
suitable for choosing the Bunch-Davies vacuum in the subhorizon
limit. At this stage, we wish to mention when do the fluctuations of
singleton become classical. The commutators in (\ref{comm}) commute
on  the superhorizon region of $z<1$ after horizon crossing.

  We are ready to compute the
power spectrum of the dipole pair defined by
\begin{eqnarray}\label{power}
_{\rm BD}\langle0|\hat{\varphi}_{a}(\eta,{\bf
x})\hat{\varphi}_{b}(\eta,{\bf y})|0\rangle_{\rm BD}=\int
d^3k\frac{{\cal P}_{\rm ab}}{4\pi k^3}e^{i{\bf k}\cdot({\bf x}-{\bf
y})}.
\end{eqnarray}
Here we choose the BD vacuum $|0\rangle_{\rm BD}$ by imposing
$\hat{c}_a({\bf k})|0\rangle_{\rm BD}=0$. On the other hand, the
cosmological  correlator defined in momentum space are  related to
the power spectra as~\cite{Baumann:2009ds}
\begin{equation} \label{mom-corr}
\langle\phi^a_{\bf k}\phi^b_{{\bf k}'}\rangle =(2\pi)^3\delta^3({\bf
k}+{\bf k}')\frac{2\pi^2}{k^3}P_{ab}(k).
\end{equation}
 Since the singleton
 theory is quite different from the two-free scalar theory, we explain what the BD vacuum is. For this purpose, we remind the reader that the
Gupta-Bleuler condition of $B^+({\bf x})|$phys$\rangle=0$ where
$B$ is a conjugate momentum of scalar photon $A_0$ was introduced
to extract the physical states of transverse photons $A_1$ and
$A_2$ by confining scalar photon $A_0$ and longitudinal photon
$A_3$ as members of quartet~\cite{AI,Kugo:1979gm}. For this
purpose, we note that the dipole pair ($\varphi_1,\varphi_2$) is
turned into the zero-norm state by making use of the BRST
transformation in Minkowski spacetime~\cite{Kim:2013mfa}. We
suggest that if the dS/LCFT correspondence works, the boundary
logarithmic operator $\sigma_2$ is related to the negative-norm
state of $\varphi_2$. In order to remove the negative-norm state,
we impose the subsidiary condition as $\varphi_1^+({\bf
x})|$phys$\rangle=0$ where $\varphi_1^+({\bf x})$ is the
positive-frequency part of the field operator. Then, the physical
space ($|$phys$\rangle$) will not include any $\varphi_2$-particle
state. This corresponds to the dipole mechanism to cancel the
negative-norm state. Here, the subsidiary condition of
$\varphi_1^+({\bf x})|$phys$\rangle=0$ is translated into
$\hat{c}_1({\bf k})|$phys$\rangle=0$ which shares a property of
the BD vacuum $|0\rangle_{\rm BD}$ defined by $\hat{c}_1({\bf
k})|0\rangle_{\rm BD}=0$, in addition to $\hat{c}_2({\bf
k})|0\rangle_{\rm BD}=0$.

The tensor power spectrum for $\varphi_1$ is given as
\begin{eqnarray}
{\cal P}_{\rm 11}=0
\end{eqnarray}
when one used  the unconventional commutation relation
$[\hat{c}_1({\bf k}), \hat{c}^{\dagger}_1({\bf k}')]=0$.

 On the other hand, it turns out that the power spectrum of $\varphi_{2}$
is defined by
\begin{eqnarray}\label{pw22}
{\cal P}_{\rm 22}&\equiv& {\cal P}_{\rm 22}^{(1)}+{\cal P}_{\rm
22}^{(2)}\nonumber\\
&=&\frac{k^3}{2\pi^2}\Bigg(\Big|\phi_{\bf k}^1\Big|^2+i(\phi_{\bf
k}^1\phi_{\bf k}^{2*}-\phi_{\bf k}^2\phi_{\bf k}^{1*})\Bigg),
\end{eqnarray}
where ${\cal P}_{\rm 22}^{(1,2)}$ denote the (first, second) term in
(\ref{pw22}) and we fixed $\tilde{N}=1/\sqrt{2k}$. Note that ${\cal
P}_{\rm 22}^{(1)}$ can be written as
\begin{eqnarray}
{\cal P}_{\rm 22}^{(1)}=\frac{k^3}{2\pi^2}\Big|\phi_{\bf k}^1\Big|^2
=\frac{H^2}{8\pi}z^3|e^{i(\frac{\pi
\nu}{2}+\frac{\pi}{4})}H_{\nu}^{(1)}(z)|^2.
\end{eqnarray}
In the superhorizon limit of $z\to 0$, the power spectrum takes the
form
\begin{equation}
{\cal P}_{\rm 22}^{(1)}\Big|_{z\to
0}=\Big(\frac{H}{2\pi}\Big)^2\Big(\frac{\Gamma(\nu)}{\Gamma(3/2)}\Big)^2\Big(\frac{z}{2}\Big)^{2w}\equiv
\xi^2z^{2w},~\xi^2=\frac{1}{2^{2w}}\Big(\frac{H}{2\pi}\Big)^2\Big(\frac{\Gamma(\nu)}{\Gamma(3/2)}\Big)^2.
\end{equation}
which implies that ${\cal P}_{\rm 22}^{(1)}$ approaches zero as
$z\to 0$. In the massless case of $m^2=0~(\nu=3/2,w=0)$, ${\cal
P}_{\rm 22}^{(1)}$ leads to the power spectrum ${\cal P}_{\rm \delta
\phi}=(H/2\pi)^2$ in (\ref{powerst}) for a massless scalar.

 It is important to note that in the superhorizon
limit of $z\to 0$, ${\cal P}_{\rm 22}^{(2)}$ is given by
\begin{eqnarray}\label{pw222}
{\cal P}_{\rm 22}^{(2)}\sim  2\xi^2 z^{2w}\ln[z],
\end{eqnarray}
which implies that ${\cal P}_{\rm 22}^{(2)}$ approaches zero as
$z\to 0$. In deriving (\ref{pw222}), $\xi$ denotes a real quantity
given by $\phi_{\bf k}^1= -i\xi z^{w}$ and $\phi_{\bf k}^2\sim \xi
z^{w}\ln[z]$.
  We mention that the remaining power spectra ${\cal P}_{\rm
12}$ and ${\cal P}_{\rm 21}$ take the same form as ${\cal P}_{\rm
22}^{(1)}$
\begin{eqnarray}
{\cal P}_{\rm 12}~=~{\cal P}_{\rm
21}&=&\frac{k^3}{2\pi^2}\Big|\phi_{\bf k}^1\Big|^2\nonumber\\
&=&{\cal P}_{\rm 22}^{(1)},
\end{eqnarray}
where we fixed $N=1/\sqrt{2k}$.

Finally, we obtain the power spectra of singleton  in the
superhorizon limit of $z\to 0$
\begin{eqnarray} \label{ps-mat1}
{\cal P}_{{ab},0}(z)&\sim&\xi^2 \left(
   \begin{array}{cc}
     0 & z^{2w} \\
     z^{2w} & z^{2w}(1+2\ln [z])\\
   \end{array}
 \right).
  \end{eqnarray}
  Its explicit form is given by
  \begin{eqnarray}
 \label{ps-mat2}
 {\cal P}_{{ab},0}(k,\eta)&\sim&\xi^2 \left(
   \begin{array}{cc}
     0 & (-k\eta)^{2w} \\
     (-k\eta)^{2w} & (-k\eta)^{2w}(1+2\ln [-k\eta])\\
   \end{array}
 \right).
 \end{eqnarray}
For $\eta=-\epsilon(0<\epsilon\ll1)$ near
$\eta=0^-$~\cite{Larsen:2003pf}, (\ref{ps-mat2}) takes the form
\begin{equation}
\label{ps-mat3}{\cal P}_{{ab},0}(k,-\epsilon)\sim \xi^2 \left(
   \begin{array}{cc}
     0 & (\epsilon k)^{2w} \\
     (\epsilon k)^{2w} & ( \epsilon k)^{2w}(1+2\ln [\epsilon k])\\
   \end{array}
 \right).
 \end{equation}
Interestingly, $k^{-3}{\cal P}_{{ab},0}(k,-1)$ has the same form as
the momentum correlators of LCFT $\langle
\sigma_a(k)\sigma_b(-k)\rangle$ with $D=(2w-1)(w-3)/(w(2w-3))$ in
(\ref{m0})-(\ref{m2}). This may show how the dS/LCFT correspondence
works for deriving the power spectra in the superhorizon limit. For
a light singleton with $m^2 \ll H^2$, one has $w\simeq
\frac{m^2}{3H^2}$. Hence, these power spectra are given by
\begin{equation} \label{ps-light}
{\cal P}_{{ab},0}|_{\frac{m^2}{ H^2}\ll1} (k,-\epsilon)\propto\left(
   \begin{array}{cc}
     0 & (\epsilon k)^{\frac{2m^2}{3H^2}} \\
     (\epsilon k)^{\frac{2m^2}{3H^2}} & (\epsilon k)^{\frac{2m^2}{3H^2}}(1+2\ln[\epsilon k])\\
   \end{array}
 \right)
 \end{equation}
 whose spectral indices are given by
\begin{equation} \label{sp-light}
n_{{ab},0}|_{\frac{m^2}{ H^2}\ll1}(k,-\epsilon)-1 =\frac{d\ln{\cal
P}_{{ab},0}|_{\frac{m^2}{ H^2}\ll1}(k,-\epsilon)}{d\ln k}= \left(
   \begin{array}{cc}
     0 & \frac{2m^2}{3H^2} \\
    \frac{2m^2}{3H^2} & \frac{2m^2}{3H^2}+\frac{2}{(1+2\ln [\epsilon k])}\\
   \end{array}
 \right).
 \end{equation}
 We observe here that $n_{{ab},0}|_{\frac{m^2}{ H^2}\ll1}$ gets a
 new contribution $\frac{2}{(1+2\ln [\epsilon k])}$ from the due to the
 logarithmic short distance singularity. Also, we observe that ${\cal P}_{{22},0}|_{\frac{m^2}{ H^2}\ll1} (k,-\epsilon) <0$ for $\epsilon k<0.607$.
 There is no such condition
 for a massive scalar propagating  on the dS spacetime.

At this stage, we briefly mention how to resolve the
$\epsilon$-dependence.  To compute the power spectra and spectral
indices correctly, one has to choose a proper slice near $\eta=0^-$.
This may be done by taking $\eta=-\epsilon$ firstly, and letting
$\epsilon \to 0$ on later. We note that the $\epsilon$-dependence
appears in the power spectra (\ref{ps-mat3}) and spectral indices
(\ref{sp-light}). As was shown in the dS/CFT
correspondence~\cite{Larsen:2003pf}, the cut-off $\epsilon$ acts
like a renormalization scale which is well-known from the UV CFT
renormalization theory.  The cosmic evolution can be seen as a
reversed renormalization group flow, from the IR fixed point (big
bang) of the dual CFT to the UV fixed point (late times) of the dual
CFT theory~\cite{Schalm:2012pi}.  Inflation occurs at a certain
intermediate stage during the renormalization group flow. This is
called as  dS holography.  Accordingly, in order to obtain the
$\epsilon$-independent power spectra and spectral indices, we should
introduce  proper counter terms to renormalize the power spectra and
spectral indices.

In the massless singleton of $m^2=0(\nu=3/2,w=0)$, the corresponding
power spectra take the form
\begin{equation} \label{ps-massless}
{\cal P}_{{ab},0}\Big|_{m^2\to 0} =\Big(\frac{H}{2\pi}\Big)^2\left(
   \begin{array}{cc}
     0 & 1 \\
     1 & 1+2\ln [z]\\
   \end{array}
 \right)
 \end{equation}
 in the superhorizon limit. Here, we note that ${\cal P}_{{12},0}|_{m^2\to 0}$ is just  the power spectrum of a
 massless scalar ${\cal P}_{\delta \phi,0}$ (\ref{fpowerst}) in the
 superhorizon limit.

\section{Discussions}

In this work, we have obtained the power spectra of singleton
generated during the dS inflation. Even though we did not know a
complete solution of $\phi^{2}_{\bf k}$ to the degenerate
fourth-order equation (\ref{s2-eq2}) in whole region, we have
obtained the power spectra which show that the dS/LCFT
correspondence plays an important role in determining the power
spectra in the superhorizon limit. Considering (\ref{mom-corr}) and
(\ref{ps-mat2}), one has  $k^{-3}{\cal P}_{{ab},0}(k,-1)\propto
\langle \phi_{\bf k}^a\phi_{-{\bf k}}^b\rangle$. Hence,  the
cosmological correlators $\langle \phi_{\bf k}^a\phi_{-{\bf
k}}^b\rangle$ are directly proportional to  the momentum
LCFT-correlators $\langle \sigma_a(k)\sigma_b(-k)\rangle$ in
(\ref{m0})-(\ref{m2}). Here we note that LCFT correlators were
derived from the ``extrapolate" dictionary (b). This is compared to
the ``differentiate" dictionary where (\ref{dir-rel})  states that
the cosmological correlator was inversely proportional to the CFT
correlator~\cite{Maldacena:2002vr}.
 Furthermore, we have
computed the spectral indices (\ref{sp-light}) for a light singleton
which contains a logarithmic correction, in compared to the massive
scalar.

In computing the power spectra, we have used  two vacua located at
$z=\infty$ (${\rm \partial dS}_\infty$) and $z=0$ (${\rm \partial
dS}_0$): the BD vacuum $|0\rangle_{\rm BD}$ in the subhorizon limit
of $z\to \infty(\eta\to -\infty)$ and the CFT vacuum $|0\rangle_{\rm
C}$ to define the correlators of operators $\sigma_a$ in the
superhorizon limit of $z\to0(\eta \to 0^-)$. The BD vacuum
$|0\rangle_{\rm BD}$ is annihilated by the two lowering operators as
$c_a({\bf k})|0\rangle_{\rm BD}=0$, and it relates to the $|{\rm
phys}\rangle$ which annihilates the negative norm state in the
quantum electrodynamics.  This is because the singleton theory is
not a two-free scalar theory. In addition, the commutation relations
(\ref{scft}) designed for the singleton quantization played an
important role to derive the power spectra in the superhorizon
limit.   On the other hand,  the CFT vacuum $|0\rangle_{\rm C}$ was
defined by imposing the Virasoro operators $L_n|0\rangle_{\rm C}=0$
for $n=0,\pm1$.  The highest-weight state $|\Phi\rangle_{\rm
C}=\Phi(0)|0\rangle_{\rm C}$ for any primary field $\Phi$ of
conformal weight $h$ is defined by $L_0|\Phi\rangle_{\rm
C}=h|\Phi\rangle_{\rm C}$ and $L_n|\Phi\rangle_{\rm C}=0$ for $n>0$.

Consequently, we have derived the power spectra and spectral indices
of singleton in the superhorizon limit by using two boundary
conditions at the infinite past ($\eta=-\infty$) and infinite future
($\eta=0^-$) where the BD vacuum was taken on the former time, while
the CFT vacuum was employed on the latter time. The dS/LCFT
correspondence was firstly realized as the computation of singleton
power spectra. Since the LCFT as dual to the singleton suffers from
the non-unitarity (for example, ${\cal P}_{{22},0}|_{\frac{m^2}{
H^2}\ll1} (k,-\epsilon) <0$ for $\epsilon k<0.607$), a truncation
mechanism will be  introduced to cure the non-unitarity in dS
spacetime~\cite{Bergshoeff:2012sc,Grumiller:2013at,Kim:2013mfa}.
 However, there remains nothing ($\sigma_{11}=0$)
for the rank-2 LCFT dual to singleton after truncating
(\ref{lcft-mat}). If one considers three-coupled scalar theory
instead of singleton, its dual correlators  will be not a $2\times
2$ matrix (\ref{lcft-mat}) but a $3\times 3$ matrix of
\begin{equation} \label{3by3}\tilde{\sigma}_{ab} \propto \left(
   \begin{array}{ccc}
     0 & 0& {\rm CFT} \\
     0 & {\rm CFT}& {\rm LCFT}\\
     {\rm CFT}&{\rm LCFT}&{\rm LCFT}^2 \\
   \end{array}
 \right).
 \end{equation}
The truncation process be carried out by throwing all terms  which
generate the third column and row of (\ref{3by3}). Actually, this
corresponds to finding a unitary CFT. We point out that a unitary
CFT ($\tilde{\sigma}_{22}$) obtained after truncation is nothing but
an ordinary CFT.

Finally, let us ask how could this scenario  account for
 cosmological observables like the amplitude of the power spectrum and the tensor-to-scalar ratio
in the cosmic microwave background. In this work, we have chosen the
dS inflation with $\dot{\phi_1}=\dot{\phi_2}=0$  instead of the
slow-roll (dS-like) inflation for simplicity. If we choose the
slow-roll inflation, then the Einstein equation takes the form of
$G_{\mu\nu}=T_{\mu\nu}/M^2_{\rm P}$ which provides the energy
density
$\rho=\dot{\phi_1}\dot{\phi_2}+(m^2\phi_1\phi_2+\mu^2\phi_1^2/2)$
and the pressure
$p=\dot{\phi_1}\dot{\phi_2}-(m^2\phi_1\phi_2+\mu^2\phi_1^2/2)$. The
first and second Friedmann equations are given by
$H^2=\frac{\rho}{3M^2_{\rm P}}$ and
$\dot{H}=-\frac{\rho+p}{2M^2_{\rm P}}$.  Also, their scalar
equations are given by $\ddot{\phi}_1+3H\dot{\phi}_1+m^2\phi_1=0$
and $\ddot{\phi}_1+3H\dot{\phi}_1+m^2\phi_2=-\mu^2\phi_1$ which are
combined to give $(\frac{d^2}{dt^2}+3H\frac{d}{dt}+m^2)^2\phi_2=0$.
However, it requires a formidable task to perform its cosmological
perturbations around the slow-roll inflation instead of the dS
inflation. Hence, we wish to remain  ``cosmological perturbations of
singleton" as a future work by answering to the question  how could
this theory account for the observed cosmological parameters in the
cosmic microwave background.

On the other hand, one may consider the holographic inflation and
thus, the dS/CFT correspondence determines the tensor central
charge. If one accepts holographic inflation such that the dS
inflation era of our universe is approximately described by a dual
CFT$_3$ living on the spatial slice at the end of inflation, the
BICEP2 results might determine the central charge $c_{\rm
T}=1.2\times 10^{9}$ of the CFT$_3$~\cite{Larsen:2014wpa}. This is
because every CFT$_3$ has a transverse-traceless tensor $T_{ij}$
with two DOF which satisfies $\langle T_{ij}({\bf x})T_{kl}({\bf
0})\rangle=\frac{c_{\rm T}}{|{\bf x}|^6} I_{ij,kl}({\bf x})$. Since
a single complex scalar $\psi$ represents two polarization modes of
the graviton, its tensor correlator in momentum space is defined by
$\langle\psi_{\bf k}\psi_{{\bf k}'}\rangle=(2\pi)^3\delta^3({\bf
k}+{\bf k}')\frac{2\pi^2}{k^3}\frac{{\cal P}_{\rm T}}{2}$ which
determines the tenor power spectrum ${\cal P}_{\rm T}=2\Big(\frac{H
t_{\rm P}}{\pi}\Big)^2={\cal P}_{{\rm h},0}$ in (\ref{fpowerst}).
This was determined to be $5\times 10^{-10}$ by
BICEP2~\cite{Ade:2014xna}. Also, its improvement of energy-momentum
tensor was reported in~\cite{Kawai:2014vxa} by including a curvature
coupling of $\zeta \phi^2 R$. As a result, if one uses the critical
gravity including curvature squared terms to describe the
holographic  inflation, the dS/LCFT picture for tensor modes would
play a role in determining other cosmological observables.

 \vspace{0.25cm}

\section*{Appendix: LCFT correlators from ``extrapolate" dictionary}

In this appendix, we derive the LCFT correlators by making use of
the extrapolation approach (b) in the superhorizon limit. For this purpose,
we consider the Green's function for a massive scalar propagating on
dS spacetime
\begin{equation} \label{green}
G_0(\eta,{\bf x};\eta',{\bf
y})=\frac{H^2}{16\pi}\Gamma(\triangle_+)\Gamma(\triangle_-)~_2F_1(\triangle_+,\triangle_-,2;1-\frac{\xi}{4})
\end{equation}
with $\xi=\frac{-(\eta-\eta')^2+|{\bf x}-{\bf y}|^2}{\eta \eta'}$.
Taking  a transformation form of hypergeometric function
\begin{eqnarray}
_2F_1(\triangle_+,\triangle_-,2;1-\frac{\xi}{4})=
\Big(\frac{4}{\xi}\Big)^{\triangle_-}~
_2F_1\Big(\triangle_-,2-\triangle_+,2;\frac{1-\frac{\xi}{4}}{-\frac{\xi}{4}}\Big),
\end{eqnarray}
we  obtain the  asymptotic form for $\triangle_-=w$
\begin{equation}\label{g0e}
\lim_{\eta,\eta'\to 0}(\eta\eta')^{-w}G_0(\eta,{\bf x};\eta',{\bf
y})\propto\frac{1}{|{\bf x}-{\bf y}|^{2w}},
\end{equation}
which corresponds to LCFT correlators $_{\rm e}\langle {\cal O}_{1}({\bf x}){\cal O}_{2}({\bf y})\rangle_{\rm e}
=_{\rm e}\langle {\cal O}_{2}({\bf x}){\cal
O}_{1}({\bf y})\rangle_{\rm e}$. Furthermore, the Green's function $G_1$ is derived by
taking derivative with respect to $w$ as
\begin{eqnarray}
G_1=\frac{d}{dw}G_0=\Big(\frac{4}{\xi}\Big)^{w}\Big(-\ln\Big[\frac{\xi}{4}\Big]+\frac{1}{F}\frac{\partial F}{\partial w}\Big)F,
\end{eqnarray}
where $F$ denotes
$F=H^2\Gamma(3-w)\Gamma(w)_2F_1(w,w-1,2;1-4/\xi)/(16\pi)$. It turns
out that its asymptotic form is given by
\begin{eqnarray}
\lim_{\eta,\eta'\to 0}(\eta\eta')^{-w}G_1(\eta,{\bf
x};\eta',{\bf y})\propto\frac{1}{|{\bf x}-{\bf
y}|^{2w}}\Big(-2\ln|{\bf x}-{\bf
y}|+\zeta_1\Big),\label{g1e}
\end{eqnarray}
where $\zeta_1$ is some constant and (\ref{g1e}) corresponds to $_{\rm e}\langle {\cal O}_{2}({\bf x}){\cal O}_{2}({\bf
y})\rangle_{\rm e}$.

\vspace{0.5cm}
 {\bf Acknowledgement}

\vspace{0.25cm}
 This work was supported by the National
Research Foundation of Korea (NRF) grant funded by the Korea
government (MEST) (No.2012-R1A1A2A10040499).

\end{document}